\documentclass[journal=jacsat,manuscript=article]{achemso}

\usepackage[utf8]{inputenc}
\usepackage[T1]{fontenc}
\usepackage{mathptmx}
\usepackage{xcolor}
\usepackage[version=3]{mhchem} 

\author{Bruno H. S. Mendon\c{c}a}
\affiliation[UFMG]{Departamento de F{\'i}sica, ICEX, Universidade Federal de Minas Gerais, CP 702, Belo Horizonte 30123-970, MG, Brazil}
\email{brunnohennrique13@gmail.com}

\author{Elizane E. de Moraes}
\affiliation[UFBA]{Instituto de F{\'i}sica, Universidade Federal da Bahia,  Campus Universit{\'a}rio de Ondina, Salvador 40210-340, BA, Brazil}

\author{Alexsandro Kirch}
\affiliation[USP]{Instituto de F{\'i}sica, Universidade de S{\~a}o Paulo, CP 66318, S{\~a}o Paulo 05315-970, SP, Brazil}

\author{Ronaldo J. C. Batista}
\affiliation[UFOP]{Departamento de F{\'i}sica, Universidade Federal de Ouro Preto, Campus Morro do Cruzeiro, Ouro Preto 35400-000, MG, Brazil}

\author{Alan B. de Oliveira}
\affiliation[UFOP]{Departamento de F{\'i}sica, Universidade Federal de Ouro Preto, Campus Morro do Cruzeiro, Ouro Preto 35400-000, MG, Brazil}

\author{Marcia C. Barbosa}
\affiliation[UFRGS]{Instituto de F{\'i}sica, Universidade Federal do Rio Grande do Sul, Porto Alegre 91501-970, RS, Brazil}

\author{H{\'e}lio Chacham}
\affiliation[UFMG]{Departamento de F{\'i}sica, ICEX, Universidade Federal de Minas Gerais, CP 702, Belo Horizonte 30123-970, MG, Brazil}

\title[An \textsf{achemso} demo]
  {Flow through deformed carbon nanotubes predicted by rigid and flexible water models}

\abbreviations{IR,NMR,UV}

\begin{document}







\begin{abstract}
In this study, using non-equilibrium molecular dynamics simulation, the flow of water in deformed carbon nanotubes is studied for two water models TIP4P/2005 and SPC/FH. The results demonstrated a non-uniform dependence of the flow on the tube deformation and the flexibility imposed on the water molecules, leading to an unexpected increase in the flow in some cases. The effects of tube diameter and pressure gradient are investigated to explain the abnormal flow behavior with different degrees of structural deformation.
\end{abstract}

\section{Introduction}

Under spatial confinement, water exhibits unusual properties, and its flow velocity may be a thousand times faster than that predicted by hydrodynamics~\cite{@10.1038/43844a,@10.1126/science.1126298,@10.1021/nl200843g,@nature/35102535,@10.1021/jp402141f}. This characteristic is desirable for applications such as drug delivery~\cite{@10.1038/nnano.2006.175,@10.1073/pnas.1004714107} and biomimetic selective ionic transport~\cite{@10.1021/nl0346326}. Also, the ionic selectivity for water desalination~\cite{@10.1166/jctn.2014.3488,@10.1021/jp709845u} and energy storage~\cite{@10.1126/science.1127261}, where the enhanced flow rates would improve the device efficiency, accuracy, and throughput.  

The large surface area-to-volume ratio inherent to the nanoscale plays a central role in the water transport properties~\cite{@10.1126/science.1126298}. From the experimental perspective, Qin et al.~\cite{@10.1021/nl200843g} observed that water velocity in carbon nanotubes (CNTs) decreases with increasing tube diameter, so the obtained values ranged between 46 (1.6~nm) and 928$\mu$~m/s (0.8~nm). Also, Secchi et al.~\cite{@10.1038/nature19315} evaluated the fluid flow rate in nanochannels and observed that the slip length increases monotonically with the decrease of tube diameter. Compared to boron nitride tubes with similar diameters, the slip length observed in CNTs was approximately 100 times higher. 

From the molecular modeling perspective, the water transport in nanotubes~\cite{@10.1006/jcph.1995.1039} was investigated with ab initio and classical molecular dynamics (MD). While the ab initio approach is limited by the number of atoms in the system, the classical results depend on reliable molecular models. In the case of water, various molecular models are available to reproduce the competing effects of hydrogen bonding and van der Waals interactions, which are responsible for water's anomalous properties~\cite{@10.1007/978-94-015-7658-1_21,@10.1080/00268978700100141,@10.1063/1.2121687,@10.1021/jp003919d,@10.1063/1.1683075,@10.1063/1.445869,@10.1063/1.481505,@10.1063/1.2056539,@10.1063/1.2907845,@10.1063/1.3124184,@10.1103/PhysRevLett.126.236001,@10.1080/00268979100102391,@10.1039/D0NR02511A}. The rigid TIP4P/2005~\cite{@10.1063/1.2121687} and the flexible SPC/FH~\cite{@10.1080/00268978700100141}  are the most popular models with reliable results describing bulk properties. However, their transferability to the confined environment is currently under discussion, and comparative studies involving different molecular models may guide performance evaluation. 

Most computer-based studies on fluid transport analyzed perfect cylindrical CNTs, and few works considered a more realistic description by including structural deformations~\cite{@10.1016/j.carbon.2015.09.099,@10.1016/j.commatsci.2004.02.018}. As typical, the synthesis process may result in defected CNTs with vacancies and structural distortions~\cite{@10.1039/C5SC04218F,@10.1021/acs.jctc.5b00292} resulted from adsorbed functional groups or mechanical compression. Structural deformations may influence the flow velocity,  shear stress, and effective viscosity~\cite{@10.1063/1.3651158,@nature19315,@10.1103/PhysRevFluids.1.054103}. Further studies on this topic may contribute to a more comprehensive understanding of mass transport in synthesized CNTs. 

We performed a systematic study to investigate the effect of structurally deformed CNTs on water transport properties within the classical non-equilibrium molecular dynamics simulations approach. The water flow rate was related to the tube chirality, diameter, and deformation level. We compared the results from rigid and flexible water models to evaluate their transferability to a confined environment context. 

The remaining of this manuscript is organized as follows. In Sec. II we present the methodology used in this analysis and the simulated models are defined. In Sec. III the discussion of the results is exposed and in Sec. IV we present the conclusions.

\section{Model and Method}

\begin{figure}[H]
\begin{center}
\includegraphics[width=6.4in]{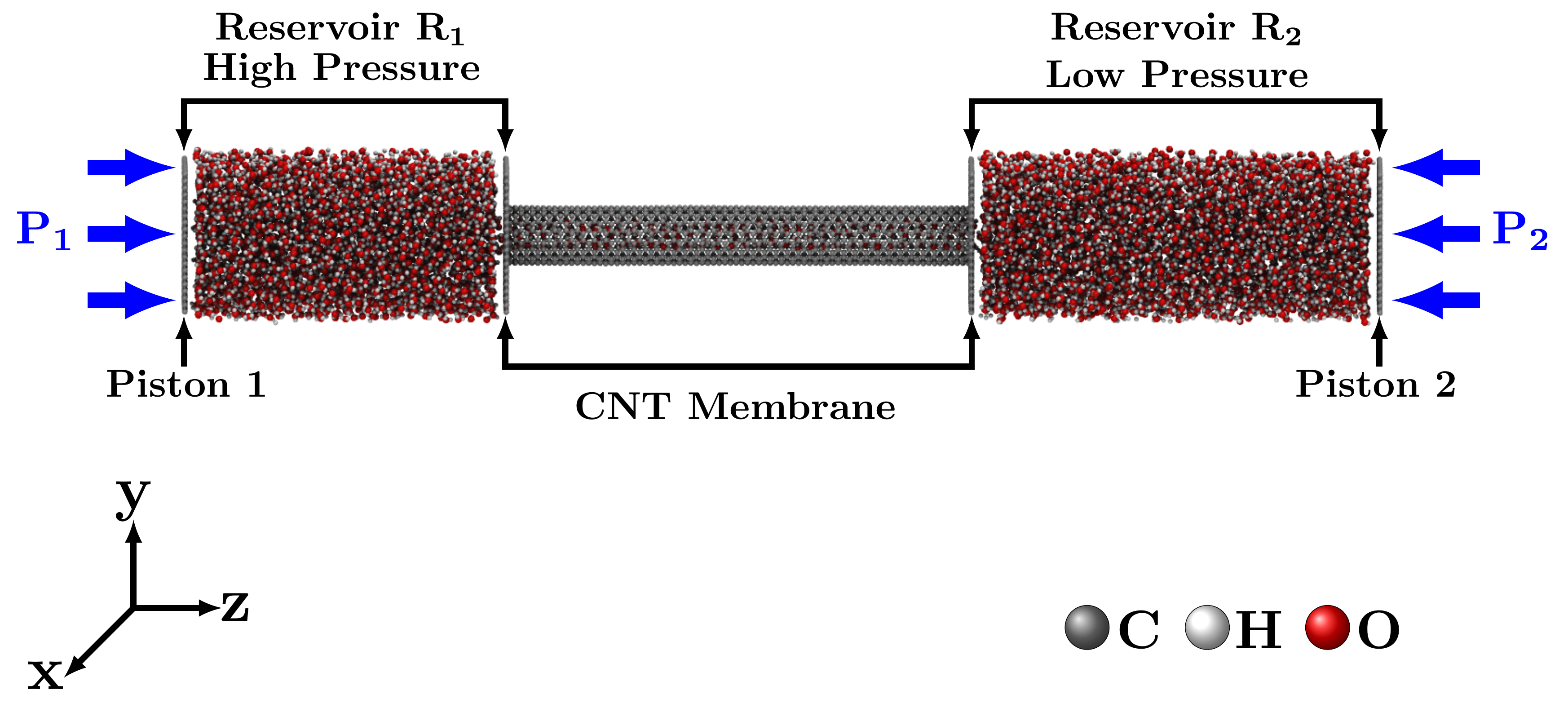}
\end{center}
\caption{Snapshot of the model system composed of two reservoirs, R$_{1}$ and R$_{2}$, with water at different pressures and limited by graphene membranes of side lengths  $L_{x}=L_{y}=3~nm$ are connected to a CNT  with length $L_{z}=10~nm$ and diameter $d$}
\label{fig_system}
\end{figure}

Our system setup is similar to the molecular model proposed by Huang et al.~\cite{@10.1063/1.2209236} where a nanotube connects two fluid reservoirs and enables the investigation of the fluid transport. The system is shown in Figure~\ref{fig_system}, and it is described as follows. Two reservoirs with water, R$_{1}$ and R$_{2}$, at different pressures, limited by two graphene membranes with size $L_{x}=L_{y}=3~nm$  are connected to a CNT with size length $L_{z}=10~nm$ and diameter $d$. The pressure at the two reservoirs is determined by two pistons, $P_1$ and $P_2$.

Outer graphene sheets on each reservoir play the role of pistons 1 and 2. The pressure on each piston is given by:

\begin{eqnarray}
\label{eq:fluxo}
P=\frac{F\cdot~n}{A} \;,
\end{eqnarray}

\noindent where $F$ is the external force in the $z$ direction applied to each atom in the graphene sheet, $n$ is the number of carbon atoms, and $A$ is the surface area. While the pressure in piston 1 ranges from 200 to 800~bar, piston 2 has a fixed value of 1 bar to impose a pressure gradient on the system.   

\begin{figure}[H]
	\begin{center}
		\includegraphics[width=6.4in]{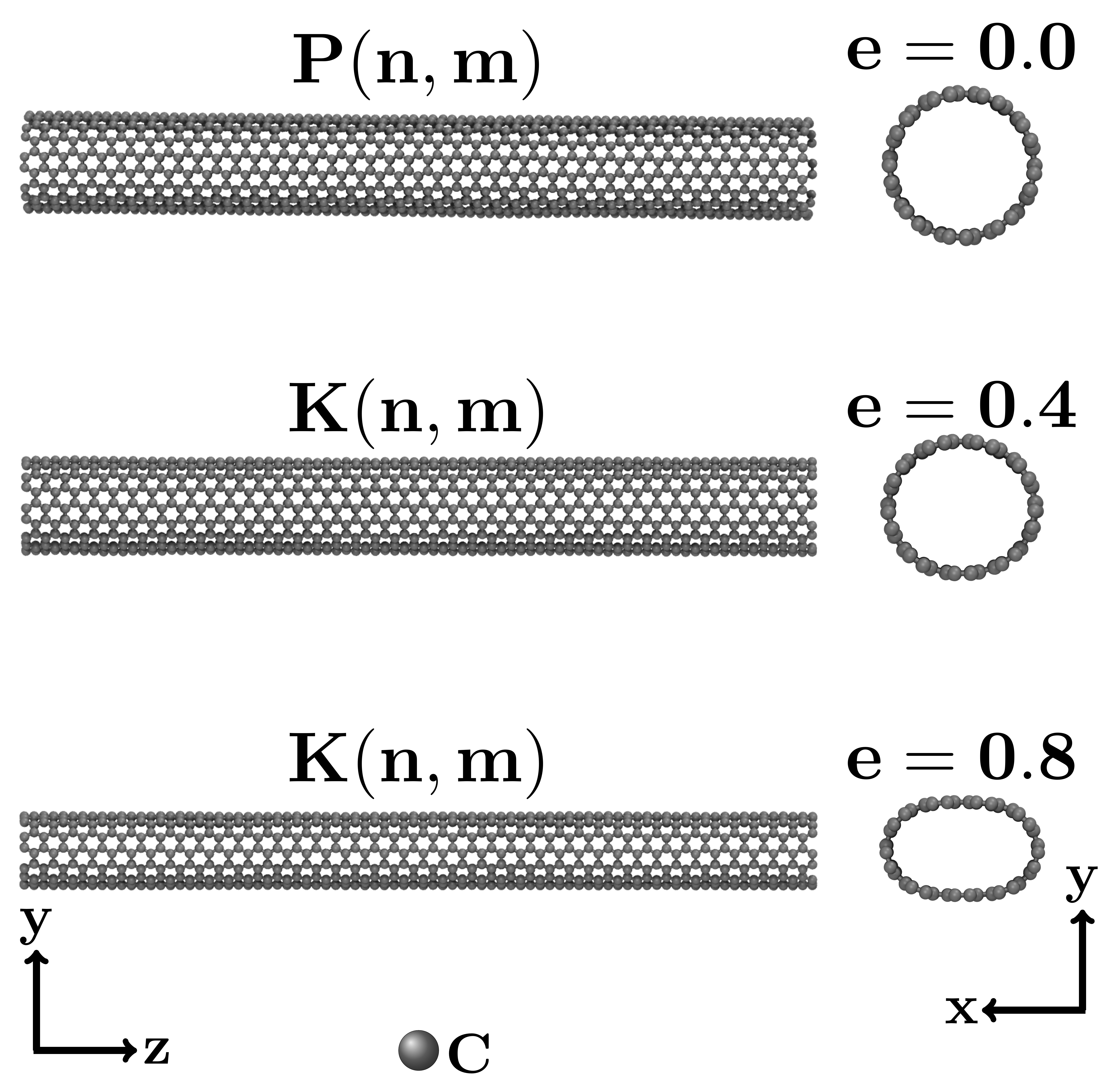}
	\end{center}
	\caption{Snapshot of the perfect P($n,m$) simulated carbon nanotubes with eccentricity equal to $0.0$, and kneaded K($n,m$) with eccentricities equal to $0.4$ and $0.8$.} 
	\label{fig_cnts}
\end{figure}

Here we study  the armchair and zigzag nanotubes with the diameters shown in Table~\ref{tab_CNT}, with armchair and zigzag displaying similar diameters for comparison. Our study considers two types of CNTs, i.e., perfect P($n,m$) and kneaded K($n,m$), as illustrated in Figure~\ref{fig_cnts}. The kneaded nanotubes K($n,m$) are produced by uniformly deforming a perfect nanotube P($n,m$)  in the radial direction until the nanotube reached an elliptical cylindrical shape with eccentricities equal to 0.4 and 0.8.  
\begin{table}[H]
\begin{center}
  \caption{Parameters for the simulated perfect carbon nanotubes ($e=0.0$) armchairs and zigzags.}
  \begin{tabular}{ c c c c c }
    \hline
    CNT    $\quad$ & $\quad$ $d$ (nm) \\ \hline
    (7,7)  $\quad$ & $\quad$ 0.95    \\
    (12,0) $\quad$ & $\quad$ 0.94    \\
    (9,9)  $\quad$ & $\quad$ 1.22     \\
    (16,0) $\quad$ & $\quad$ 1.25     \\
    (12,12)$\quad$ & $\quad$ 1.63     \\
    (21,0) $\quad$ & $\quad$ 1.64     \\ 
    \hline 
  \end{tabular}
\label{tab_CNT}
\end{center}
\end{table}

The water molecules are represented by two different models: the flexible SPC/FH~\cite{@10.1063/1.3124184}, and the rigid TIP4P/2005~\cite{@10.1063/1.2121687}.  The parameters considered for the water models were defined and are shown in the Table~\ref{tab_watermodels}. In this water models, the Lennard-Jones site is located on the oxygen atom, with parameters $\sigma$ and $\epsilon$. The charges of oxygen and hydrogen are $q_{O}$ and $q_{H}$, respectively. The TIP4P/2005 model places a negative charge $q_{M}$ at a point M at a distance $d_{OM}$ from the oxygen along the H-O-H bisector. The distance between the oxygen and hydrogen sites is $r_{OH}$. The angle formed between hydrogen, oxygen and another hydrogen atom is given by $\theta_{HOH}$. For flexible model SPC/FH, the $k_{OH}$ and $k_{\theta}$ are the potential depth parameters, and OH and $\theta$ are the reference bond length and angle, respectively.

The SHAKE~\cite{@10.1016/0021-9991(77)90098-5} algorithm is employed to stabilize the molecule bonds and angles. The nanotube and the graphene sheets are modeled by the Lennard-Jones potential (LJ) considering fixed bond lengths and angles~\cite{@10.1088/0959-5309} with effective carbon-carbon interaction energy $\epsilon_{CC}=0.086$~kcal$\cdot$mol$^{-1}$ and an effective diameter of $\sigma_{CC}=3.4$~\AA~\cite{@nature/35102535}. The carbon-oxygen energy $\epsilon_{CO}=0.11831$~kcal$\cdot$mol$^{-1}$ and the effective carbon-oxygen diameter $\sigma_{CO}=3.28218$~\AA~\cite{@nature/35102535}. The Lorentz-Berteloth mixing rules provided the LJ crossing parameters.
\begin{table}[H]
	\begin{center}
	\caption{Force field parameters used for each of the water models. }
		\begin{tabular}{ c c c }
			\hline
			  & TIP4P/2005 & SPC/FH   \\ \hline
                $\epsilon_{OO}$ (kcal mol$^{-1}$) & 0.1852  & 0.1553 \\
                $\epsilon_{HH}$ (kcal mol$^{-1}$) & 0.0   & 0.0396 \\
                $\sigma_{OO}$ (\AA) & 3.1589  & 3.188 \\
                $\sigma_{HH}$ (\AA) & 0.0  & 0.65 \\	
                $q_{O}$ (e) & 0.0  & -0.8476 \\	
                $q_{H}$ (e) & 0.5564  & 0.4138 \\	
                $q_{M}$ (e) & -1.1128  & * \\
                $d_{OM}$ (\AA) & 0.1546 & * \\	
                $r_{OH}$ (\AA) & 0.9572  & 1.0 \\		
                $\theta_{HOH}$ ($^{\circ}$) & 104.52  & 109.4 \\
                $k_{OH}$ (kcal mol$^{-1}$ \AA$^{-2}$) & *  & 1108.580 \\	
                $k_{\theta}$ (kcal mol$^{-1}$ rad$^{-2}$) & *  & 91.53 \\	      	
                \hline 
		\end{tabular}
		\label{tab_watermodels}
	\end{center}
\end{table}

We adopted the Large-scale Atomic/Molecular Massively Parallel Simulator (LAMMPS)~\cite{@10.1006/jcph.1995.1039}  package  to perform the simulations, with the Particle-Particle Particle-Mesh (PPPM) method to calculate long-range Coulomb interactions~\cite{@10.1021/acs.jpcc.7b08326}. 

The system's dynamics features were evaluated by considering the flow rate calculations as given by the equation:

\begin{eqnarray}
\label{eq:MSD}
\phi_{H_{2}O}=A\;v \;,
\end{eqnarray}

\noindent where $A$ is the graphene layer area ($34$x$34$~\AA$^{2}$), and $v$ is the water flow velocity acquired from the least-square linear regression line fitted to the data cloud which relates the average molecular displacement along the tube axis as a function of the time as taken from the MD trajectory file.

The simulation protocol involves the following steps:

\begin{enumerate}
    \item Pre-equilibrium in the NVE ensemble with a 0.5 ns  MD run to minimize system energy keeping the pistons frozen (net force equal to zero). 
    \item Forces are applied in the pistons in order to impose 1 bar in each system to reach the water equilibrium densities at 300 K. Equilibration in the NPT ensemble during 1.0 ns. 
    \item  Pistons are freezed in the new equilibrium position. Equilibration in the NVT ensemble at 300 K controlled via the Nosé-Hoover thermostat~\cite{@10.1080/00268978400101201} during 2.0 ns. 
    \item Nanopores are opened. Different forces are applied in each piston to mimic the pressure gradient. NPT ensemble during 10 ns at 300 K and different feed pressures.
\end{enumerate}

\section{Results and discussion}

Water transport in carbon nanotubes is highly dependent on the diameter of the tubes~\cite{10.1098@rsta.2015.0357,10.1016@j.desal.2015.02.040,10.1039@C8CP01191E,@10.1016/j.physa.2018.11.042,@10.1063/1.5129394,@10.1063/5.0031084,@10.1016/j.chemphys.2020.110849}. Considering the high resistance to flow in the inlet and outlet regions of small diameter CNTs, the flow rate, and water molecule flux through the CNTs will decrease markedly with decreasing CNT diameters~\cite{10.1063@1.5000493}. For small diameter tubes, the size of the fluid particles becomes crucial and occupies a considerable portion in determining mobility. But for those with a larger diameter, these effects will diminish~\cite{10.1063@1.4948071,10.3390@nano10061203}. Some experimental data significantly suggest that the configuration of water in nanotubes can vary according to their diameters~\cite{@10.1016/j.physa.2018.11.042,@10.1063/1.5129394,@10.1063/5.0031084,@10.1016/j.chemphys.2020.110849,10.1021@jacs.6b02635,10.1016@j.ces.2018.05.018}.

Even though the flow decreases with the decrease of the diameter, the ratio between the flow obtained in the simulation and the value classical hydrodynamic predicts for diameters below 1 nm, increases with the decrease of the diameter because in this regime the water flows in a stressless single line. This enhancement flow with the decrease of the diameter is larger for armchair than for the zigzag nanotubes~\cite{sam2019}. This indicates that the spiral-like path of the water at armchair nanotubes exhibits less stress than the ring-like move of water at the zigzag nanotube. Below we test how this is affected by the deformation and by the nature of the water model.

\begin{figure}[H]
	\begin{center}
		\includegraphics[width=6.4in]{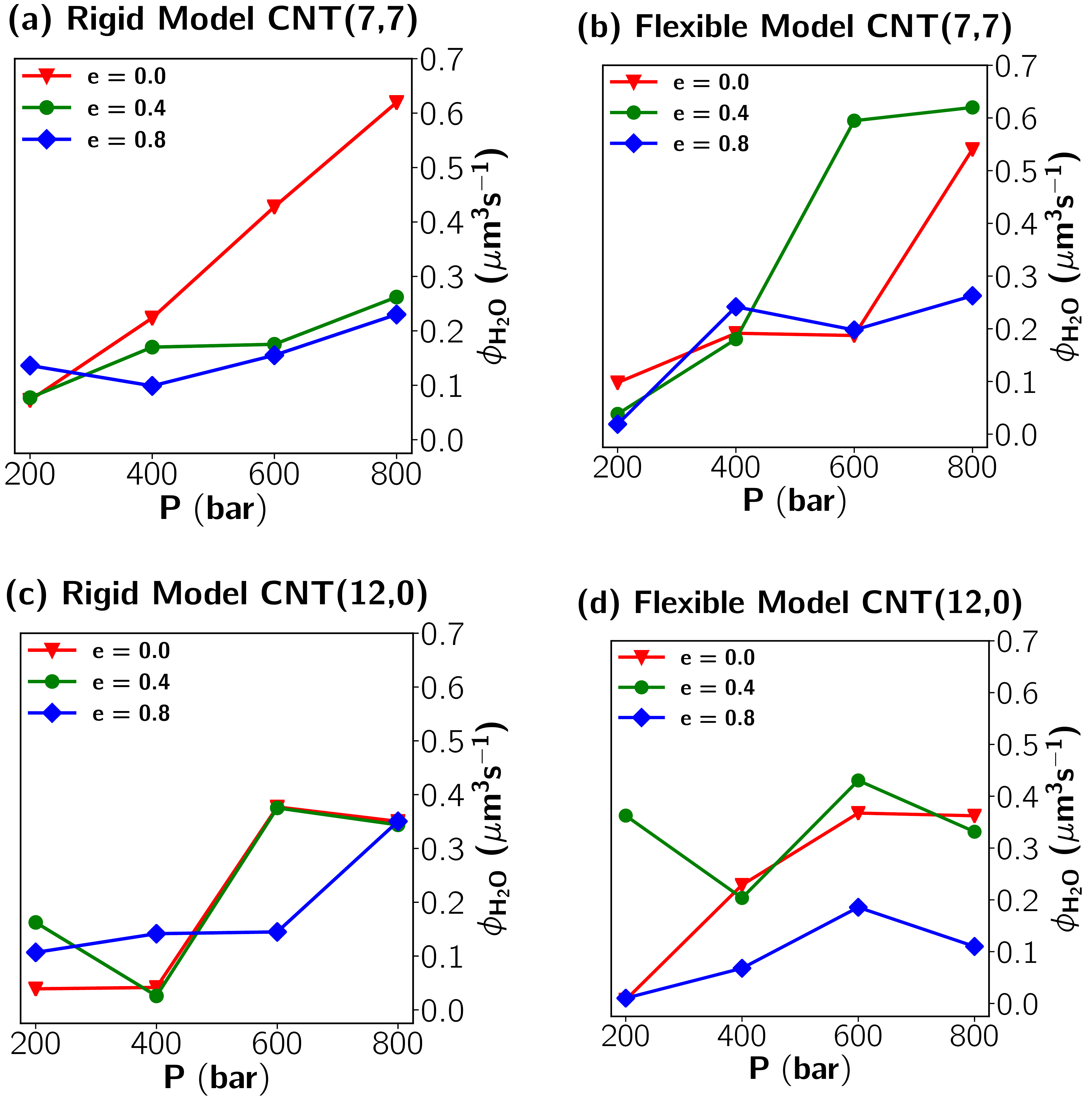}
	\end{center}
	\caption{Flow rate as a function of the pressure gradient applied to the carbon nanotubes, ($7,7$) and ($12,0$), perfect P($n,m$) with $e=0.0$ and kneaded K($n,m$) with $e=0.4$ and $0.8$ for rigid TIP4P/2005 and flexible SPC/FH water models.}
 \label{fig_7x7-12x0}
\end{figure}

Figure~\ref{fig_7x7-12x0} (a) shows the flow rate as a function of pressure for the armchair (7,7) nanotube for the TIP4P/2005 model for the prefect and two cases of kneaded nanotube. The behavior with pressure is linear for all the three cases. The nanotube deformation does not change the linear regime but decreases the flow, particularly for larger pressures. This decrease can be attributed to the introduction of additional stress due to deformation. Figure~\ref{fig_7x7-12x0} (b) also shows the flow rate versus pressure, but for the SPC/FH model the small deformation increases the flow rate, possibly because the hydrogen bonds adapt to enhance mobility. Larger deformations lead to an increase of stress and the mobility decreases.

Figure~\ref{fig_7x7-12x0} (c) illustrates the flow rate versus pressure for the (12,0) zigzag nanotube for the TIP4P/2005 model for the prefect and two cases of kneaded nanotube. As observed in the perfect nanotube case~\cite{sam2019} the flow for the water in armchair is larger than for the zigzag nanotubes. The small deformation does not affect the flow rate. The same behavior is observed for the flexible SPC/FH shown in the Figure~\ref{fig_7x7-12x0} (d).

This result seems to reinforce the assumption that the large enhancement factor observed for water in armchair nanotubes is due to the spiral path the water molecules perform. Any deformation, which disrupts the spiral path, decreases the flow in the case of the armchair nanotubes.

\begin{figure}[H]
	\begin{center}
		\includegraphics[width=6.4in]{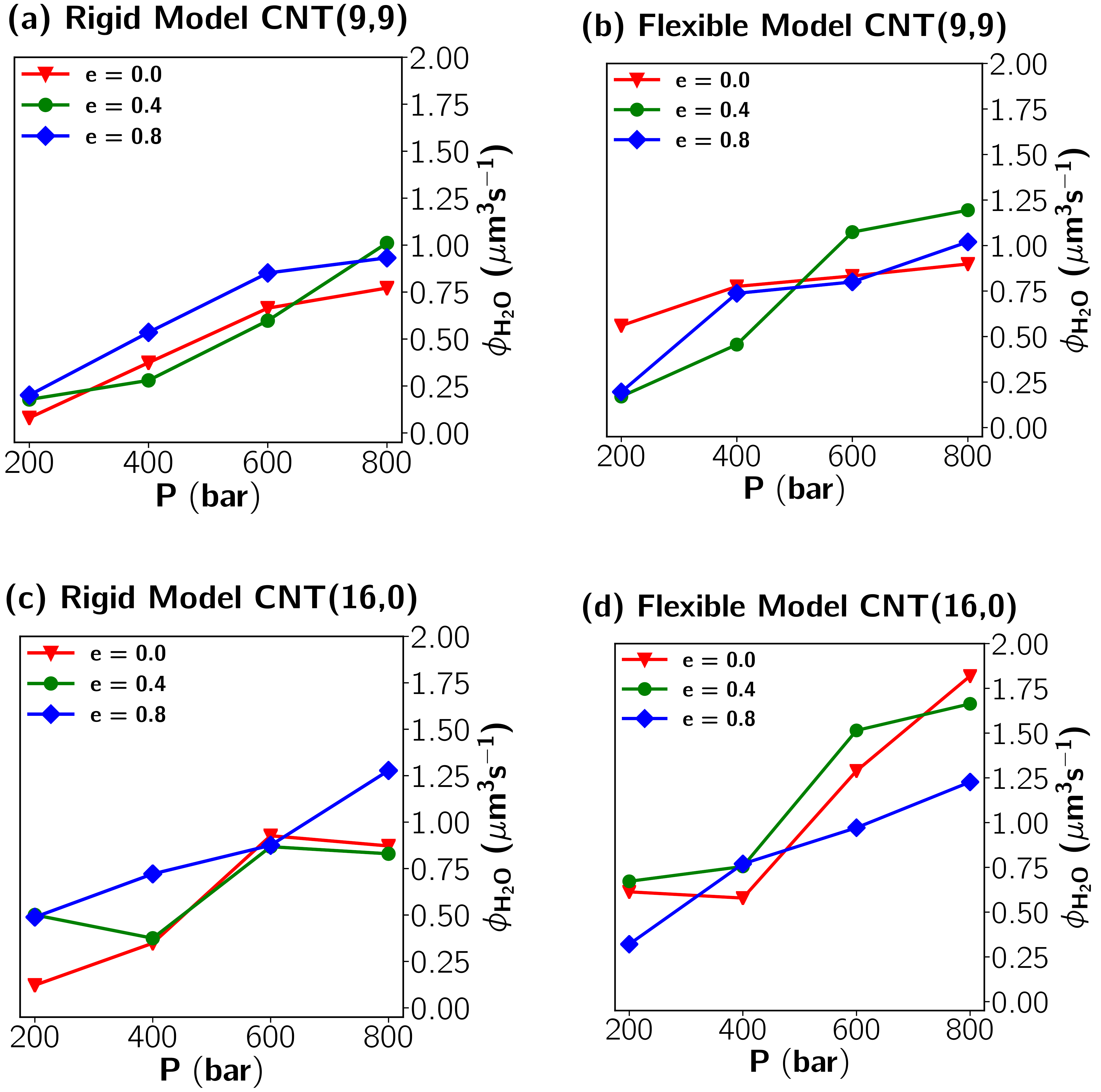}
	\end{center}
	\caption{Flow rate as a function of the pressure gradient applied to the carbon nanotubes, ($9,9$) and ($16,0$), perfect P($n,m$) with $e=0.0$ and kneaded K($n,m$) with $e=0.4$ and $0.8$ for rigid TIP4P/2005 and flexible SPC/FH water models.} 
	\label{fig_9x9-16x0}
\end{figure}

In the case of the (16,0) zigzag nanotube the flow, illustrated in 
Figure~\ref{fig_9x9-16x0}(a) and (b) shows the water flux versus pressure for an armchair ($9,9$) nanotube for the TIP4P/2005 and SPC/FH models for the prefect and two cases of kneaded nanotube. In this diameter, both the diffusion~\cite{@10.1063/1.5129394} and the enhancement factor~\cite{@10.1021/nl200843g} show a decrease when compared with smaller and larger nanotube diameters, what is attributed to the transition between a single line of water at smaller diameters to a single spiral of water molecules. In the case of the flow,  both TIP4P/2005 and SPC/FH are strongly affected by the deformation. The SPC/FH due to flexibility of the hydrogen bonds show a larger mobility when compared with the rigid TIP4P/2005 model.

Figure~\ref{fig_9x9-16x0}(c) and (d) for the  TIP4P/2005 and SPC/FH models show an increase in water mobility when compared with the equivalent armchair nanotube. Even though in general water is more mobile in armchair nanotubes~\cite{sam2019}, results for diffusion~\cite{@10.1006/jcph.1995.1039} show that the ring-like structure observed in the (16,0) is more mobile than the spiral-like organization seen in the (9,9) case. The larger mobility observed for the flexible model is not surprising. It indicates a better accommodation of the hydrogen bonds.

\begin{figure}[H]
	\begin{center}
		\includegraphics[width=6.4in]{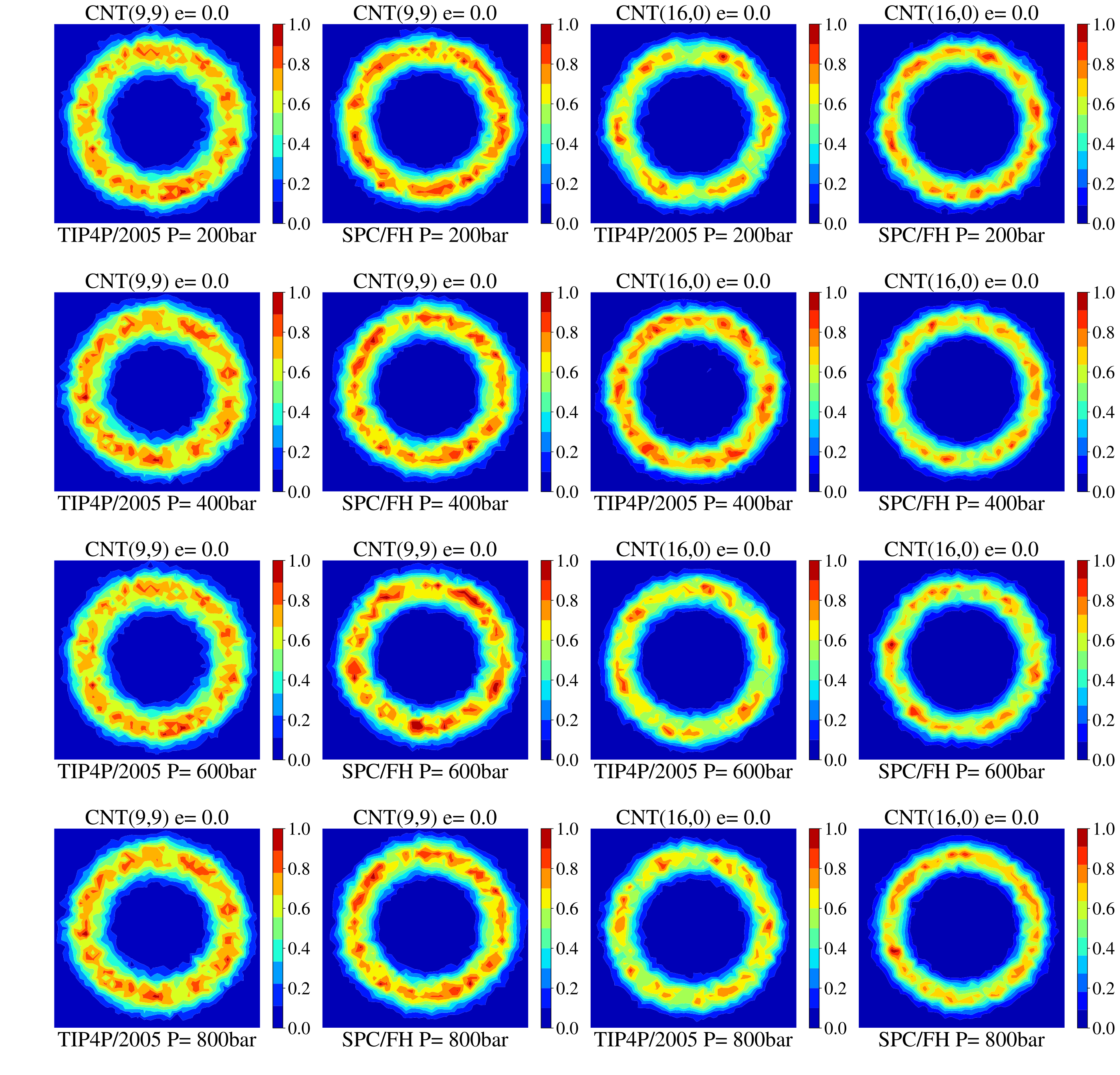}
	\end{center}
	\caption{Density maps in the $xy$ direction for the carbon nanotubes ($9,9$) and ($16,0$) perfect P($n,m$), and the comparison of the rigid TIP4P/2005 and flexible SPC/FH water models. Dark blue regions have a low probability of finding water molecules, while red regions have a high probability of finding water molecules.}     
	\label{fig_maps-9x9-16x0-P}
\end{figure}

For the perfect CNT ($9,9$) and ($16,0$) illustrated in Figure~\ref{fig_maps-9x9-16x0-P} water forms similar structures regardless of the wall. The two models of water adapt to the applied pressures forming the same structural arrangement in both cases, causing the water to move with the same structure. An increase in the pressure applied to the systems would favor the formation process of hydrogen bonds and, therefore, the effect of the nanotube structure would be predominant in the water flow. This phenomenon explains the close water flow found for these diameters.

\begin{figure}[H]
	\begin{center}
		\includegraphics[width=6.4in]{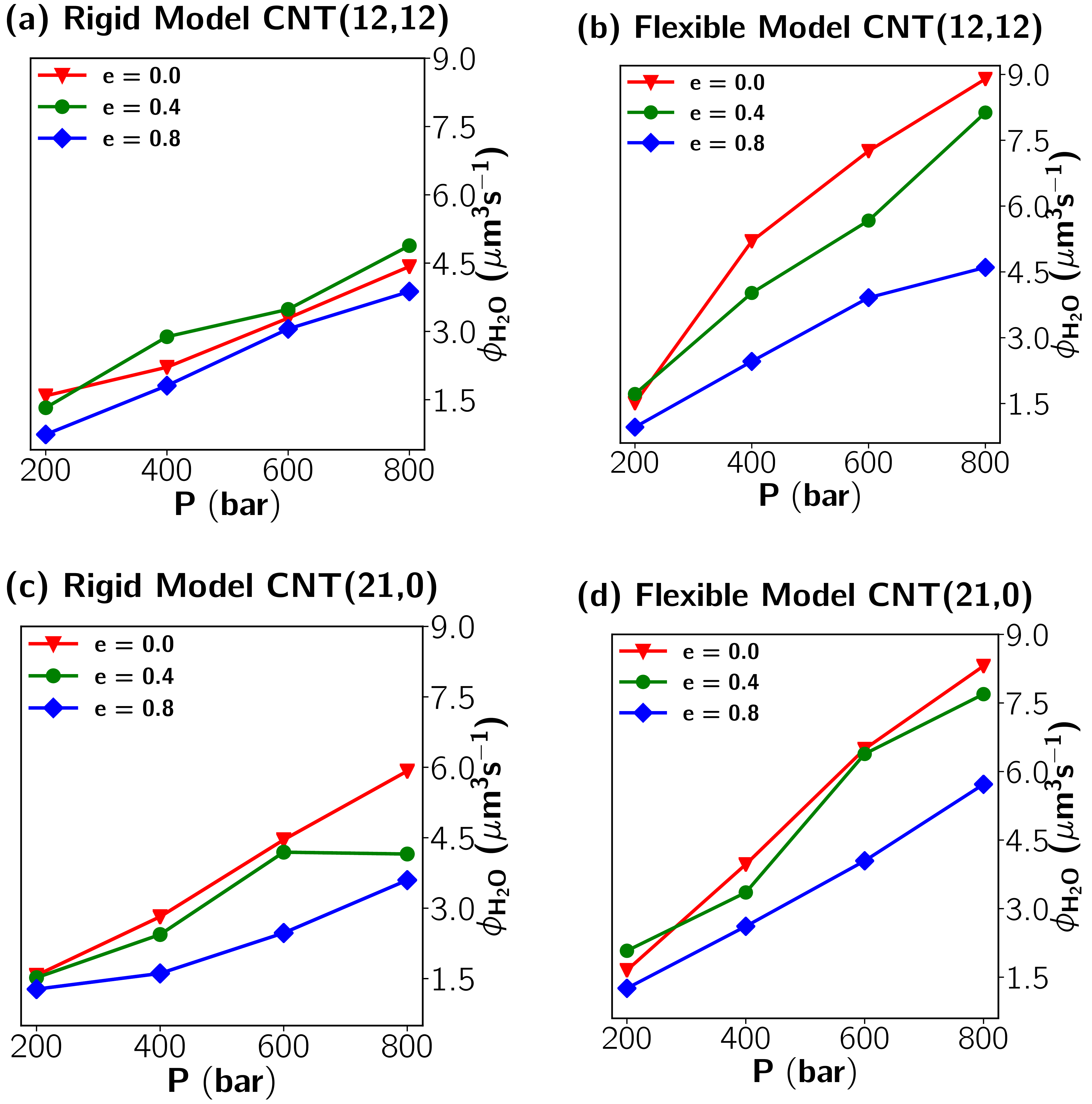}
	\end{center}
	\caption{Flow rate as a function of the pressure gradient applied to the carbon nanotubes, ($12,12$) and ($21,0$), perfect P($n,m$) with $e=0.0$ and kneaded K($n,m$) with $e=0.4$ and $0.8$ for rigid TIP4P/2005 and flexible SPC/FH water models.} 
	\label{fig_12x12-21x0}
\end{figure}

For CNT($12,12$) and CNT($21,0$) carbon nanotubes shown in Figure~\ref{fig_12x12-21x0}, as the pressure increases, rapid progress of water flow can be observed for both water models. The improvement in water flow as a function of the diameter of carbon nanotubes and the imposition of flexibility on water molecules is remarkable. By increasing the diameter of the cylindrical pore section, matching the membranes with the CNTs ($12,12$) and ($21,0$), the increase in water flux notably increases almost twice for the SPC/FH compared to the TIP4P/2005, indicating a strong dependence not only on the membrane diameter but also on the water model. For the SPC/FH model, this expressive increase in flow is due to the difference in connection and angle length and to the characteristic flexibility of the model. Another fact that explains this phenomenon is the increase in the space available for water molecules to pass each other. At larger CNT diameters, fluid-fluid interactions became more important as the CNT internal surface area to fluid volume ratio decreased.

\section{Conclusions}
We investigated the influence of deformations on carbon nanotubes with different chiralities and subjected to different pressure gradients for the transport of confined water, helping to understand the relevance of the water model and these defects in improving the flow. We considered the TIP4P/2005 and SPC/FH models for the study. 

The simulations showed how deformations on carbon nanotubes cross-section affect the internal flow dynamics. The inclusion of deformations in tubes of smaller diameters, ($7,7$) and ($12,0$) tend to considerably reduce the fluid velocity for larger deformations. The flow rate is further affected by a reduction in the overall density of the fluid caused by the presence of deformations. For large deformations, $e=0.8$, the inclusion of strains disrupts the smooth, continuous potential energy landscape that a CNT provides, causing greater pressure losses along it, and reducing the overall fluid velocity. 

The overall observation indicates that flow enhancement factors exhibit a considerable reduction when subjected to substantial degrees of deformation. However, there are specific cases in which tubes with diameters measuring $12.2$ and $12.4$ Angstroms demonstrate an opposite trend, with flow rates tending to increase as the level of deformation rises. Water transport is not strongly affected by CNT chirality for different diameters and strains for both models, except for CNTs ($9,9$) and ($16,0$). This diameter is quite different when compared to smaller and larger nanotubes because only a layer of water close to the wall is formed. This layer of water behaves quite differently depending on the chirality. The higher value deformations, in general, showed much less flow when compared to the perfect nanotube. The number of molecules passing through the pore gradually decreases with an increasing degree of deformation. For some cases, the flow does not show any significant variation. When analyzing the rate at which water molecules pass through the nanochannels, the values found for the different diameters suggest that carbon nanotube membranes apparently can be competitive systems with currently existing water filtration processes. 

These results indicate that carbon nanotube membranes are promising for use as filters for impure water or saline water, although more studies are needed to categorically affirm their viability. Furthermore, advances in experimental measurements of nanoconfined water will lead to more accurate models of water in carbon nanotube simulations. Currently, atomistic water models can provide a reasonable range of results for water flowing through CNTs. Understanding the impacts of simulation choices aids in the analysis of past results and improves the design of future simulation studies.

\begin{acknowledgement}
This work is funded by the Brazilian scientific agency Conselho Nacional de Desenvolvimento Científico e Tecnológico (CNPq) and the Brazilian Institute of Science and Technology (INCT) in Carbon Nanomaterials with collaboration and computational support from  Universidade Federal de Minas Gerais (UFMG), Universidade Federal de Ouro Preto (UFOP), Universidade Federal da Bahia (UFBA), and Universidade de São Paulo (USP). ABO thanks the Brazilian science agency FAPEMIG for financial support. BHSM wish to acknowledge João P. K. Abal (UFRGS) for discussions and theoretical contributions. In  addition, the authors acknowledge the National Laboratory for Scientific Computing (LNCC/MCTI, Brazil) for providing HPC resources of the SDumont supercomputer, which have contributed to the research results reported within this paper. URL: http://sdumont.lncc.br.
\end{acknowledgement}

\bibliography{achemso-demo}

\end{document}